\documentclass[11pt,a4paper]{article}
\usepackage{amsmath,amssymb}
\usepackage{epsfig,graphicx}
\usepackage{cite}

\begin{document}

\title{\bf Neutrino photoproduction on electron in dense magnetized medium}

\author{M.~V.~Chistyakov\footnote{{e-mail}: mch@uniyar.ac.ru},
A.~V.~Kuznetsov\footnote{{e-mail}: avkuzn@uniyar.ac.ru},
N.~V.~Mikheev\footnote{Deceased},
D.~A.~Rumyantsev\footnote{{e-mail}: rda@uniyar.ac.ru},
\\
D.~M.~Shlenev
\\
\small{\em Yaroslavl State P.~G.~Demidov University}
\\
\small{\em Sovietskaya 14, 150000 Yaroslavl, Russian Federation}
}
\date{}
\maketitle

\begin{abstract}
The effect of a strongly magnetized cold plasma on the Compton-like 
photoproduction of a 
neutrino-antineutrino pair on an electron,  $\gamma e \to e \nu \bar \nu$, has been considered. 
The contribution of this process to the neutrino 
emissivity in both the non-resonance and the resonance cases has been calculated 
with taking account of the photon dispersion 
properties in medium. 
 Our results show that the 
neutrino emissivity owing to the
$\gamma e \to e \nu \bar \nu$ 
reaction is significantly modified as compared to the previously reported 
data.

\end{abstract}

\def\mprp{\mbox{\tiny $\bot$}}
\def\mprl{\mbox{\tiny $\|$}}

\def\beq{\begin{eqnarray}}
\def\eeq{\end{eqnarray}}
\def\ee{\varepsilon}
\def\lm{\lambda}
\newcommand{\prp}[1]{#1_{\mbox{\tiny $\bot$}}}
\newcommand{\prl}[1]{#1_{\mbox{\tiny $\|$}}}
\def\ff{\Lambda}
\def\tff{\widetilde \Lambda}

\newcommand{\ii}{\mathrm{i}} %
\newcommand{\dd}{\mathrm{d}} %
\newcommand{\eee}{\mathrm{e}} %


\section {Introduction}\label{Sec:1}


The problem of a correct description of effects
of an active external medium (a strong magnetic field
and/or dense plasma) on quantum processes is of
current interest during a long time, for a review see e.g.~\cite{KM_Book_2013}.
First, these effects are caused by the sensitivity 
of charged fermions (primarily, electrons as particles 
with the largest specific charge) to the field.
Second, a strongly magnetized plasma significantly 
changes the dispersion properties of photons
and, thereby, the kinematics of the processes.

The conditions of the strongly magnetized plasma
can exist in the interiors of magnetars, i.e., isolated
neutron stars with magnetic fields much higher than
the critical value  $B_e = m^2/e \simeq 4.41\times 10^{13}$~G
~\footnote{We use the natural units
$c = \hbar = k_{\mathrm{B}} = 1$, $m$ is the electron mass, and $e>0$ is the elementary
charge.}. Recent observations (see~\cite{Olausen:2014} and the papers 
cited therein) make it possible, in particular, to
identify some astrophysical objects (SGR and AXP) as
magnetars.

At the same time, all known theoretical models of
the internal structure of neutron stars give the parameters 
of the medium (density and temperature) at
which the magnetized plasma is transparent to neutrinos. 
In this case, reactions with a neutrino-antineutrino pair 
in the final state are decisive for neutrino 
cooling. In this connection, the neutrino photoproduction (the so-called photoneutrino process)
$\gamma e \to e \nu \bar \nu$ was studied by different authors~\cite{Skobelev:2000,RCh:2008,Borisov:2011,Borisov:2012}. 
The formulas were obtained for the neutrino emissivity, i.e., the energy 
carried by the neutrino from the unit volume of a star per unit time, for both the nonrelativistic
and relativistic plasmas. However, in those papers the anisotropy in the
dispersion of photons was disregarded, which can change the results essentially. 
Furthermore, the expressions for the neutrino emissivity
owing to the Compton-like process $\gamma e \to e \nu \bar \nu$ in
nonrelativistic and relativistic plasmas contain  
some inaccuracies~\cite{RCh:2008,Borisov:2011,Borisov:2012}.

In this work, we study in detail the Compton-like
photoproduction of the neutrino-antineu\-tri\-no pair $\gamma e \to e \nu \bar \nu$ and the
neutrino emissivity owing to this process with an
accurate inclusion of the dispersion properties of 
photons  in dense magnetized medium
in both  resonance and non-resonance cases.


\section {Photon dispersion in the magnetized medium}\label{Sec:2}


The propagation of electromagnetic radiation in
any active medium is conveniently described in terms
of normal modes (eigenmodes). In turn, the 
polarization and dispersion properties of normal modes are
connected with the eigenvectors and the eigenvalues of the 
polarization operator ${\cal P}_{\alpha \beta}$, 
respectively. It is
convenient to decompose the tensor ${\cal P}_{\alpha \beta}$ in terms of
the basis of 4-vectors~\cite{BatShab:1971} constructed of the 
electromagnetic field tensor reduced to a dimensionless form,
and the 4-momentum of a photon $q^{\alpha} = (\omega, {\bf k})$:
\beq
\label{eq:basis}
b_{\mu}^{(1)} = (\varphi q)_\mu, \qquad
 b_{\mu}^{(2)} = (\tilde \varphi q)_\mu, 
\\
\nonumber
b_{\mu}^{(3)} = q^2 \, (\Lambda q)_\mu - q_\mu \, q^2_{\mbox{\tiny $\bot$}}, 
\qquad b_{\mu}^{(4)} = q_\mu, 
\eeq 
\noindent 
which are the eigenvectors of the polarization operator
in a static uniform magnetic field. In this case,  
$(b^{(1)} b^{(1)}) = -q^2_{\mprp}$, 
$(b^{(2)} b^{(2)}) = -q^2_{\mprl}$, $(b^{(3)} b^{(3)}) = -q^2 q^2_{\mprl} 
q^2_{\mprp}$, $(b^{(4)} b^{(4)}) = q^2$.

Hereafter we use the following notations:
the 4-vectors with the indices $\bot$ and $\parallel$
belong
to the Euclidean \{1, 2\} subspace and the Minkowski \{0, 3\} subspace correspondingly.
Then for arbitrary 4-vectors
$A_\mu$, $B_\mu$ one has
\beq
&&A_{\mprp}^\mu = (0, A_1, A_2, 0), \quad  A_{\mprl}^\mu = (A_0, 0, 0, A_3), \nonumber \\
&&(A B)_{\mprp} = (A \Lambda B) =  A_1 B_1 + A_2 B_2 , \nonumber \\
&&(A B)_{\mprl} = (A \widetilde \Lambda B) = A_0 B_0 - A_3 B_3, \nonumber
\end{eqnarray}

\noindent where the matrices
$\Lambda_{\mu \nu} = (\varphi \varphi)_{\mu \nu}$,\,
$\widetilde \Lambda_{\mu \nu} =
(\tilde \varphi \tilde \varphi)_{\mu \nu}$ are constructed with
the dimensionless tensor of the external
magnetic field, $\varphi_{\mu \nu} =  F_{\mu \nu} /B$,
and the dual tensor,
${\tilde \varphi}_{\mu \nu} = \frac{1}{2}
\varepsilon_{\mu \nu \rho \sigma} \varphi^{\rho \sigma}$.
The matrices $\Lambda_{\mu \nu}$ and  $\widetilde \Lambda_{\mu \nu}$
are connected by the relation
$\widetilde \Lambda_{\mu \nu} - \Lambda_{\mu \nu} =
g_{\mu \nu} = \rm{diag} (1, -1, -1, -1)$,
and play the role of the metric tensors in the perpendicular ($\bot$)
and the parallel ($\parallel$) subspaces respectively.

We emphasize that, in contrast to the magnetic
field, the value ${\cal E}_e = B_e$ for the electric field is limiting
because the generation of the nearly critical electric
field in a macroscopic region of space would result in
the intense production of electron-positron pairs
from vacuum. At the same time, if the electric field is
perpendicular to the magnetic field, the electric field ${\cal E}_e$
can exceed the critical value $B_e$, remaining below $B$.
However, even in this case, it is always possible to 
perform a Lorentz transformation to a reference frame
where only the magnetic field exists. This statement
can be generalized to the case where the plasma with a
temperature $T$ and a chemical potential $\mu$ moves as a
whole along the magnetic field. To this end, it is 
sufficient to represent the electron distribution function in
an explicit Lorentz-invariant form in terms of the 4-velocity 
of the medium $u_{\alpha}$:
\beq
\label{eq:fermidist}
f_{e}(p) = \frac{1}{1+\exp{[((pu)_{\mprl}-\mu)/T]}} \, , \quad (pu)_{\mprl} = E u_0 - p_z u_z \, , \quad 
E=\sqrt{p_z^2+m^2} \, .
\eeq

The condition of the absence of the electric field in
this frame can be written in the relativistically 
covariant form $(u \Lambda)_\mu = 0$. Consequently, when the plasma
moves as a whole along the magnetic field, it is possible
to consider the situation where only the magnetic field
exists.

In  this case  we can obtain the following expansion of 
${\cal P}_{\alpha \beta}$ over the eigenvectors $r_{\alpha}^{(\lambda)}$ 
with the eigenvalues $\varkappa^{(\lambda)}$~\cite{Rojas1979,Rojas1982,Shabad:1988,MRCh:2014}:
\beq
\label{eq:Pab1}
{\cal P}_{\alpha \beta} = \sum_{\lambda = 1}^{3} 
 \varkappa^{(\lambda)} \, \frac{r_{\alpha}^{(\lambda)} 
(r_{\beta}^{(\lambda)})^{*}}{(r^{(\lambda)})^2} \, , \quad 
r_{\beta}^{(\lambda)} = \sum\limits_{i = 1}^{3} A_i^{(\lambda)} \, b_{\beta}^{(i)} \, , 
\eeq
\noindent where  $A_i^{(\lambda)}$ are some complex coefficients.
As we can see from Eq.~(\ref{eq:Pab1}), it is rather difficult to determine 
the dispersion properties of photons for all
three polarizations, and it is true even in the strongly magnetized
plasma approximation, $\beta \gg m^2,\, \mu^2, \, T^2$ (hereafter $\beta = eB$). However, as the analysis 
shows (see, e.g.,~\cite{RCh:2008, KM_Book}), 
in the case $\beta \gg m^2$, where electrons occupy the ground Landau level, only photons
with the polarization corresponding to the vector
$r_{\alpha}^{(2)} \simeq b^{(2)}_{\alpha}$ will provide the leading contributions 
to the amplitude of the $\gamma e \to e \nu \bar \nu$ process in the
external field.
 In the cold plasma approximation, $\omega, T \ll \mu - m$,
the expression for the eigenvalue $\varkappa^{(2)}$ can be obtained analytically and represented 
in the form
\beq
\label{eq:kappa20}
\varkappa^{(2)} \simeq  \frac{\omega_p^2 \,q^2_{\mprl}}{\omega^2-v_{\mathrm{F}}^2 k_z^2} \, ,  \quad 
v_{\mathrm{F}} = \frac{p_{\mathrm{F}}}{\mu} = \sqrt{1-\frac{m^2}{\mu^2}} \, , 
\eeq
\noindent  where $\omega_p^2 = (2\alpha \beta/\pi) \, v_{\mathrm{F}}$ is the so-called plasma frequency
squared, $p_{\mathrm{F}}$ is the Fermi momentum.

In this case, the solution of the dispersion equation
\beq
q^2 - \varkappa^{(2)} = 0 \, 
\label{disper}
\eeq
\noindent for a photon of the mode 2 propagating at a nonzero angle  $\theta$
to the magnetic field direction  can be found analytically as the function
$\omega = \omega ({\bf k})$ taking the form
\beq
\label{eq:omega2}
\omega &=& \bigg \{ \frac{1}{2} \left [k^2 (1+v_{\mathrm{F}}^2 \cos^2{\theta}) + \omega_p^2 \right ] 
\\ [3mm]
\nonumber
&+& \frac{1}{2} \sqrt{\left [k^2 (1-v_{\mathrm{F}}^2 \cos^2{\theta}) + \omega_p^2 \right ]^2 - 4\omega_p^2 
(1-v_{\mathrm{F}}^2) k^2 \cos^2{\theta}} \bigg \}^{1/2} \, .
\eeq

In the case of a nonrelativistic plasma, where
$v_{\mathrm{F}} \ll 1$, we obtain
\beq
\label{eq:omega2nrel}
\omega \simeq \bigg \{ \frac{1}{2} \left [k^2 + \omega_p^2 \right ] +
\frac{1}{2} \sqrt{\left [k^2 + \omega_p^2 \right ]^2 - 4\omega_p^2 
 k^2 \cos^2{\theta}} \bigg \}^{1/2} \, .
\eeq

The corresponding expression for the relativistic
plasma, where $v_{\mathrm{F}} \to 1$, has the form
\beq
\label{eq:omega2rel}
\omega \simeq \sqrt{k^2+\omega_p^2} \, .
\eeq

One more question which appears to be important in some cases, is
the renormalization of the wavefunction of a photon. 
We note that this renormalization becomes insignificant in the case of a cold plasma because
the main contribution to the physically observed characteristics (e.g., emittance) comes from the photon
energy range $\omega \ll m$.


\section {Neutrino emissivity in the non-resonance case}\label{Sec:3}


Our main goal is to obtain the expression for the neutrino emissivity 
caused by the process $\gamma e \to e \nu \bar \nu$. 
In turn, the neutrino emissivity can be defined as the zero component of the 
four-vector of the energy-momentum carried away by the neutrino pair due to
this process from a unit volume of plasma per unit time. 
Here, we neglect the inverse effect of the energy and momentum loss on the state of plasma. 
In the conditions of a strongly magnetized plasma 
the neutrino emissivity can be represented in the form~\cite{Yakovlev2000}:
\beq
\label{eq:Q}
Q_{\gamma e \to e \nu \bar \nu} &=& \frac{1}{L_x}\;  
\int \frac{\dd^3 k}{(2\pi)^3 \, 2 \omega} \, f_\gamma (\omega) \, 
\frac{\dd^2 p }{(2\pi)^2 \, 2 E} \, 
f_e (E) \,  
\frac{\dd^2 p'}{(2\pi)^2 \, 2 E^{\, \prime} } \, \left [1-f_e (E^{\, \prime}) \right ] \,  
\\[3mm]
\nonumber
&\times&
\frac{\dd^3 p_1}{(2\pi)^3 \, 2 \varepsilon_1} \,
\frac{\dd^3 p_2}{(2\pi)^3 \, 2 \varepsilon_2} \,  q^{\, \prime}_0\,
(2 \pi)^3  \, \delta^3 (P - p^{\, \prime} - q^{\, \prime}) 
|{\cal M}_{\gamma e \to e \nu \bar \nu}|^2 \, ,
\eeq
where $f_\gamma (\omega) =  \left [\eee^{\omega/T} - 1 \right ]^{-1}$ is the equilibrium 
distribution function of an initial photon with the four-vector $q^{\mu} = (\omega, {\bf k})$, 
 $f_e (E)$ and $f_e (E^{\, \prime})$ are the equilibrium distribution functions of
initial and final electrons, respectively, in the plasma rest frame
$f_e (E) = \left [\eee^{(E - \mu)/T} + 1 \right ]^{-1}$ (see Eq.~(\ref{eq:fermidist}));  
$q^{\, \prime}_0 = \varepsilon_1 + \varepsilon_2$ is the neutrino pair energy, 
$\varepsilon_{1,2} = |{\bf p}_{1,2}|$; $\dd^2 p = \dd p_y \dd p_z$; 
$V = L_x L_y L_z$ is the plasma volume, $P_{\mu} = (p+q)_{\mu}$.

In calculating the amplitude ${\cal M}_{\gamma e \to e \nu \bar \nu}$ of the 
process $\gamma e \to e \nu \bar \nu$, 
we consider the case of relatively small momentum transfers 
compared with the $W$ boson mass, $|q^{\, \prime 2}| \ll m_W^2\,$. 
Then the corresponding interaction Lagrangian can be written as follows:
\beq
{\cal L}  =  \frac{G_{\mathrm{F}}}{\sqrt 2}\,
\big [ \bar \Psi \gamma^{\alpha} (C_V - C_A \gamma_5) \Psi \big ] \,
 j^{\, \prime}_{\alpha}  
+
 e (\bar \Psi \gamma^\alpha \Psi) \, A_\alpha \, ,  
\label{eq:Lgammanu}
\eeq
where $j^{\, \prime}_{\alpha} = \bar \nu \gamma_{\alpha} (1- \gamma_5) \nu$ is the 
current of left-handed neutrinos, $A_{\alpha}$ is the four-potential of the photon field, 
$C_V = \pm 1/2 + 2 \sin^2 \theta_W, \, C_A = \pm 1/2$, 
$\theta_\mathrm{W}$ is the Weinberg angle. 
Here, the upper sign corresponds to electron neutrinos 
($\nu = \nu_e$), when there is an exchange reaction both of $W$ and $Z$ bosons. 
The lower sign corresponds to the $\mu$ and $\tau$ neutrinos, when there is only 
$Z$ boson exchange.

\begin{figure}
\centerline{\includegraphics[width=8cm]{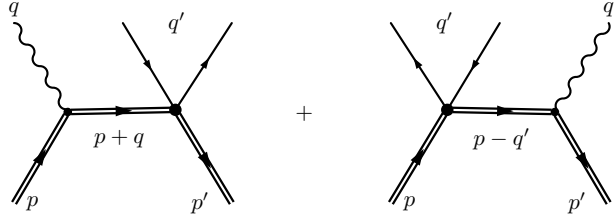} }
\caption{ The Feynman diagrams for the reaction $\gamma e \to e \nu \bar \nu$. 
Double lines mean that the effects of an external field on the initial and final electron states 
and on the electron propagator are exactly taken into account.}
\label{fig:Diag}
\end{figure}

The amplitude of the process $\gamma e \to e \nu \bar \nu$ in the tree approximation is 
described by the Feynman diagrams 
shown in Fig.~\ref{fig:Diag} and in the non-resonance case (i.e. when the virtual electron is 
on the ground Landau level) has the form~\cite{RCh:2008}
\begin{eqnarray}
\label{eq:ampl1}
{\cal M}_{\gamma e \to e \nu \bar \nu} = 
2\sqrt{2}\, e \, G_{\mathrm{F}}\, m \;
\left [C_V\, (q^{\, \prime}\tilde \varphi j^{\, \prime}) - C_A\, (q^{\, \prime} 
\tilde \varphi \tilde \varphi j^{\, \prime}) \right] 
\frac{\sqrt{q_{\mprl}^2 (|Q^2_{\mprl}|+4m^2)}}{(qq')_{\mprl}^2 - 
\varkappa^2 (q\tilde \varphi q')^2} \, ,   
\end{eqnarray}
\noindent 
where $q^{\,\prime \mu} = (p_1 + p_2)^{\mu}$ is the total
4-momentum of the neutrino-antineutrino pair, $\varkappa = \sqrt{1 - 4m^2/Q^2_{\mprl}}$,
 $Q^{\mu} = (q-q^{\,\prime})^{\mu}$.

The resulting expression for the emissivity of the
photoneutrino process can be significantly simplified
in two limiting cases.

(i) In the case of the nonrelativistic plasma, $\mu \sim m$,
and at an arbitrary relation between the
plasma frequency and the temperature, the emissivity can be presented in the form
\beq
\label{eq:Q1emu}
Q_{\gamma e \to e \nu \bar \nu} \simeq Q_{nr} \; F\left (\frac{\omega_p}{T} \right ) \, ,
\eeq
\noindent where
\beq
\label{eq:Q1Skob}
Q_{nr} = \frac{8 \pi^2 \alpha \, G_{\mathrm{F}}^2 \, \beta \, T^9}{4725 \, m \, p_{\mathrm{F}}} 
\left [\overline{C_V^2} + 
\overline{C_A^2} \right ] \simeq 1.3 \times 10^6 B_{15}^2 \, \rho_6^{-1} \, T_8^9 \left(
\frac{\mbox{erg}}{\mbox{cm}^3 \; \mbox{s}} \right)
\eeq
\noindent is the emissivity in the limit $\omega_p \ll T$~\cite{Skobelev:2000}. 
In Eq.~(\ref{eq:Q1Skob}) 
$B_{15} = B/(10^{15} \, \mbox{G})$, $\rho_{6} = \rho/(10^{6} \, \mbox{g/cm}^3)$, 
$T_{8} = T/(10^{8} \, \mbox{K})$, and the constants $\overline{C_V^2} = 0.93$ and 
$\overline{C_A^2} = 0.75$ are the results of summation over all channels of the neutrino production 
of the types $\nu_e, \nu_\mu, \nu_\tau$.

The function $F(\omega_p/T)$ depending on the ratio of the
plasma frequency to the temperature can be represented in the form of the single integral
\beq
\label{eq:Fs}
F(y) = \frac{15}{8 \pi^8}\int \limits_{y}^{\infty} \frac{\dd x \; x^5}{\eee^x - 1} 
\left (x^2-\frac{y^2}{5} \right )\, . 
\eeq
\noindent This integral can be approximated by the formula
\beq
\label{eq:FsG}
F(y) \simeq \frac{3 \eee^{-y}}{4\pi^8} \big (2 y^7 + 15 y^6 + 95 y^5 + 495 y^4 + 
2040 y^3 + 6240 y^2 + 12600 y + 12600 \big )\, .
\eeq

\begin{figure}[htb]
\centerline{\includegraphics{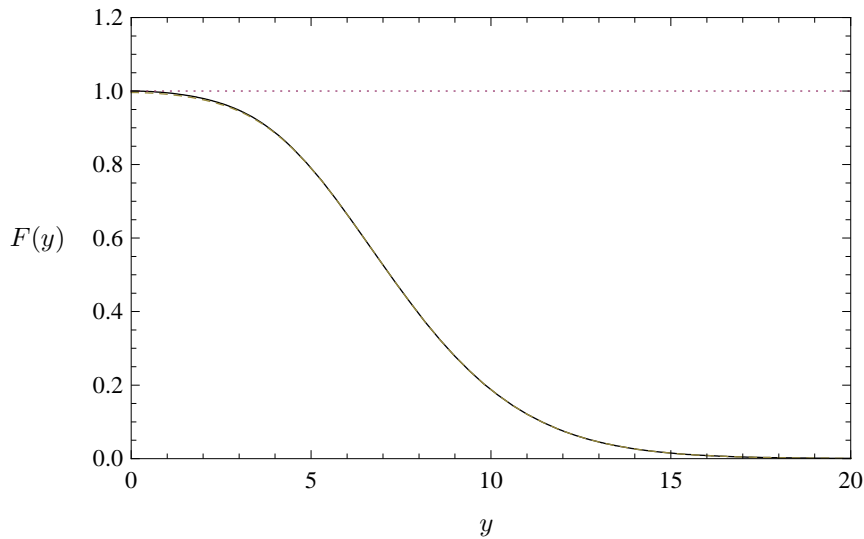}}
\caption{The function $F(y)$ defined by Eq.~(\ref{eq:Fs}). The horizontal
dotted straight line corresponds to the value $F(0) = 1$.}
\label{fig:F}
\end{figure}

The plot of the function $F(y)$ is shown in Fig.~\ref{fig:F}.
We note that the numerical analysis of the integral~(\ref{eq:Fs}) as
compared to the approximation~(\ref{eq:FsG}) gives a discrepancy no
more than $0.5\%$.

Our results~(\ref{eq:Q1emu})--(\ref{eq:FsG}) for the case of the nonrelativistic plasma  
should be compared with the results of the previous calculations, 
see Eq.~(56) of Ref.~\cite{Skobelev:2000}, Eq.~(32) of Ref.~\cite{RCh:2008} 
and Eq.~(32) of Ref.~\cite{Borisov:2012}. 
There was a discrepancy between those formulas in the numerical factor of $\pi/2$. 
Now we should confess that the factor of $\pi/2$ was lost in Eq.~(32) of Ref.~\cite{RCh:2008}.
However, it should be noted that all those formulas, with the indicated correction, were obtained 
in the limit $\omega_p \ll T$, 
i.e. when the approximation for the function~(\ref{eq:Fs}) was taken: $F(y) \to F(0) = 1$. 
As one can see from Fig.~\ref{fig:F}, this approximation is valid for the case $\omega_p \lesssim T$, 
but it would give an essentially overestimated result in the case $\omega_p > T$.
The results~(\ref{eq:Q1emu})--(\ref{eq:FsG}) obtained can be used for 
an arbitrary relation between the plasma frequency and the temperature. 

(ii) In the case of relativistic plasma, $\mu \gg m$, and at an arbitrary relation between the
plasma frequency and the temperature, the emissivity can be presented in the form

\beq
\label{eq:Q2emu}
Q_{\gamma e \to e \nu \bar \nu} \simeq Q_{r} R\left (\frac{\omega_p}{2T} \right ) \,, 
\eeq
\noindent where
\beq
\label{eq:Q2Bor}
Q_{r} &=& \frac{G_{\mathrm{F}}^2 \, \alpha \,
(\overline{C_V^2} + \overline{C_A^2})}
{576 \,(2 \pi)^{11/2} } \frac{B}{B_e} 
\left (\frac{m}{\mu} \right )^6 \omega_p^{15/2} \, T^{3/2} \,
\eee^{-\omega_p/T}
\eeq
\noindent is the emissivity in the limit $\omega_p \gg T$~\cite{Borisov:2011,Borisov:2012}.


The function $R(z)$ can be represented in the form of
the double integral
\beq
\label{eq:R}
R(z) &=& \frac{3 z^{3/2} }{5\sqrt{\pi}} \; \eee^{2z} 
\int \limits_0^{\infty} \dd v v^6  
 \eee^{-zv} \int \limits_0^1 
\frac{\dd t t^4[1 - (v-vt)^{-2}]}{1 - \eee^{-z[v(1-t) + (v-vt)^{-1}]}}\; 
\times
\\
\nonumber
&\times&\frac{vt - (v-vt)^{-1}}{1 - \eee^{-z[vt - (v-vt)^{-1}]}} 
\;[vt - 5(v-vt)^{-1}] \, ,
\eeq

\noindent which can be approximated well by the formula
\beq
\nonumber
R(z) &\simeq & 1+ \frac{0.7627}{z^{1/2}} + \frac{66.875}{z^{3/2}} + \frac{271.654}{z^{5/2}} + 
\frac{2509.36}{z^{7/2}} + \frac{6754.62}{z^{9/2}} +  
\\ [3mm]
&+&\frac{16612.9}{z^{11/2}} +\frac{19843.8}{z^{13/2}} + \frac{10188.5}{z^{15/2}} \, .
\label{eq:Rasymp}
\eeq

\begin{figure}[htb]
\centerline{\includegraphics{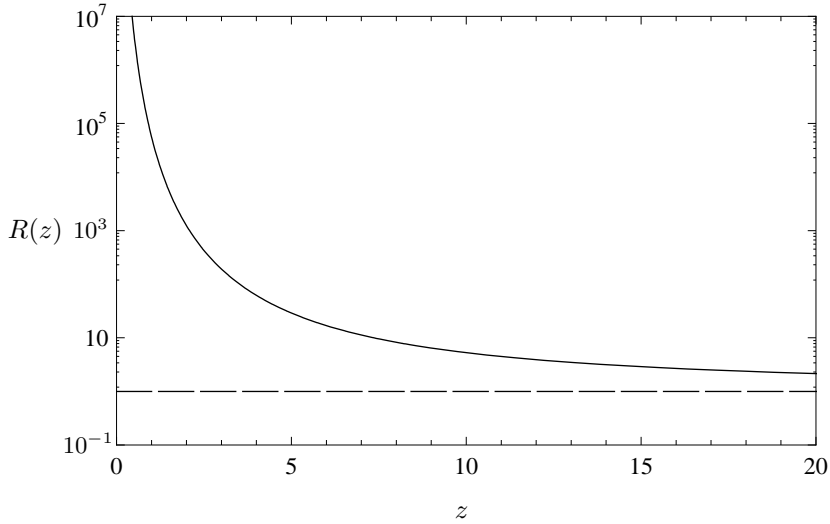}}
\caption{The function $R(z)$ defined by Eq.~(\ref{eq:R}). The horizontal
dashed straight line corresponds to the asymptotic value $R(\infty) = 1$.}
\label{fig:R}
\end{figure}

The plot of the function $R(z)$ is shown in Fig.~\ref{fig:R}. We
note that the numerical analysis of the integral~(\ref{eq:R}) as
compared to the approximation~(\ref{eq:Rasymp}) gives a discrepancy no
more than $0.8\%$. 
Therefore, the approximation~(\ref{eq:Rasymp}) can
be used for applications at an arbitrary ratio $\omega_p / T$.

According to Fig.~\ref{fig:R} and approximate formula~(\ref{eq:Rasymp}),
for the realistic parameters of the field and plasma 
 of neutron stars, $B \simeq 50 \, B_e$, $T \simeq 10^8$~K~\cite{Yakovlev2000}, the
factor is $R(14) \simeq 3$ and the asymptotic expression
$R(\infty) \simeq 1$ is inapplicable. Furthermore, it follows from
Eq.~(\ref{eq:Q2emu}) that the numerical estimate presented in Ref.~\cite{RCh:2008} is
strongly overstated. At the same time, for the parameters, used in Ref.~\cite{Borisov:2011}: 
$B \simeq 10^{16} \, \mbox{G}$, $\rho \simeq 10^{9} \, \mbox{g/cm}^3$, $T \simeq 10^{9} \, \mbox{K}$, we 
obtain from Eq.~(\ref{eq:Q2Bor}): $Q_r \simeq  10^{11} \, 
\mbox{erg}/ (\mbox{cm}^{3} \, \mbox{s})$, but for the same parameters one has 
$R(2.8) \simeq 260$. Thus, the estimate obtained in Ref.~\cite{Borisov:2011} is understated 
by several orders of magnitude.
 However, for the realistic parameters of the inner crust of neutron stars (see above)
 we obtain 
from Eq.~(\ref{eq:Q2emu}): $Q_{\gamma e \to e \nu \bar \nu} \simeq 6 \times 10^{-7} \, 
\mbox{erg}/ (\mbox{cm}^{3} \, \mbox{s})$. It is reasonable to neglect the 
neutrino photoproduction,  $\gamma e \to e \nu \bar \nu$, in the models of cooling of neutron 
stars at the matter density  $\rho \simeq 10^{9} \, \mbox{g/cm}^3$. But at the same time,  
the resonance on the virtual electron in the process $\gamma e \to e \nu \bar \nu$ 
becomes possible at the same density. Because of this, the value $Q_{\gamma e \to e \nu \bar \nu}$ 
would increase (see the next section).


\section{Resonance in the process $\gamma e \to e \nu \bar \nu$}\label{Sec:4} 


On the boundary between the outer and the inner crust of a magnetar 
(where the electron 
density is estimated as $\rho \gtrsim 10^{9} \, \mbox{g/cm}^3$), 
the higher Landau levels for a virtual electron begin to be excited. 
  In this case the denominator of the electron propagator $P^2_{\mprl} - m^2 - 2e B n$ 
 can be equal to zero and the resonance on the virtual electron become possible. 
  In addition, the virtual electron resonance occurs only in the $s$-channel 
diagram (the first diagram in Fig.~\ref{fig:Diag}). 
 On the other hand, to accurately take into account the resonance behavior
in the process $\gamma e \to e \nu \bar \nu$, it is necessary to calculate 
radiative corrections to the electron mass, caused by the combined action of a 
magnetic field and plasma. This calculation is a separate challenge. 
However, because of the smallness of these corrections, we can approximately replace 
$m^2 \to m^2 - \ii P_0 \Gamma_n$ in the denominator of  
the electron propagator, such that 
\beq
\frac{1}{P^2_{\mprl} - m^2 - 2 \beta n} \to  
\frac{1}{P^2_{\mprl} - m^2 - 2 \beta n + \ii P_0 \Gamma_n} \, .
\label{eq:denominator}
\eeq

In this case, the main contribution to the amplitude arises from the resonance region, 
so that we can approximately replace the corresponding part of the amplitude 
(see~\cite{Kuznetsov:2013}) by the
$\delta$ function:
\beq
\nonumber
&&|{\cal M}_{\gamma e \to e \nu \bar \nu}|^2 \simeq   \sum \limits_{n=1}^{\infty} \; 
 \frac{|{\cal R}_{n}|^2}{(P^2_{\mprl} - m^2 - 2\beta n)^2 + P_0^2 \Gamma_n^2}
\\
\label{eq:ampl3}
&& \simeq  \sum \limits_{n=1}^{\infty} \; 
\frac{\pi}{P_0 \Gamma_n}   \, 
\delta (P^2_{\mprl} - m^2 - 2 \beta n) \, |{\cal R}_{n}|^2 \, ,
\eeq
where $\Gamma_n$ is the total width of the change of the electron state. 
This width can be represented in the form~\cite{Weldon:1983} 
\beq
\label{eq:weldon}
\Gamma_n = \Gamma^{abs} + \Gamma^{cr} \simeq \Gamma^{cr}_{e_0 \gamma \to e_n} 
\left [1+ \eee^{(E^{\, \prime \prime}_n - \mu)/T} \right ]  \, .
\eeq
Here 
\beq
\nonumber
&&\Gamma^{cr}_{e_0 \gamma \to e_n}  = \frac{1}{2 E''_n} \, 
\int \frac{\dd^3 k }{2 \omega (2\pi)^3} \, f_\gamma (\omega) \, 
\frac{\dd^2 p }{2 E (2\pi)^2} \, f_e (E) \, 
\\
\label{eq:gammacr1}
&&\times (2\pi)^3 \, \delta^3 (P - p^{\, \prime \prime})
\, |{\cal M}_{e_0 \gamma  \to e_n}|^2  
\eeq 
is the width of the electron creation in the $n$th Landau level.

On the other hand, in the case of resonance the expression for $|R_n|^2$ being averaged over 
the photon polarizations can be factored in the strong field limit $\beta \gg m^2$
as follows (see, for example,~\cite{Latal:86}):
\beq
\label{eq:factor}
|{\cal R}_n|^2 = |{\cal M}_{e_0 \gamma  \to e_n}|^2 \, |{\cal M}_{e_n \to e_0 \nu \bar \nu}|^2 \, ,
\eeq
where 
\beq
\label{eq:factor1}
|{\cal M}_{e_0 \gamma  \to e_n}|^2 = \frac{8\pi \alpha}{n!} \, 
\exp{\left (-\frac{q^2_{\mprp}}{2\beta} \right )}
\left (\frac{q^2_{\mprp}}{2\beta} \right )^n \, M_n^2 \, (p \tilde \Lambda q)  \, 
\sum\limits_{\lambda = 1}^{3} 
\bigg [ \left |A_1^{(\lambda)} \right |^2 + 
\left |A_2^{(\lambda)} + \sigma A_3^{(\lambda)} \right |^2 \bigg ]
\eeq
is the amplitude of the absorption of a photon in the 
process $e_0 \gamma \to e_n$, when an electron passes from the ground Landau level 
to a higher Landau level $n$. 
The parameter $\sigma = (p \tilde \varphi q)/(p \tilde \Lambda q) = \pm 1$  
determines the direction of the photon propagation with respect to the magnetic field direction. 
The values $A_i^{(\lambda)}$ are introduced in Eq.~(\ref{eq:Pab1}).
Finally, the amplitude squared of the electron transition from the $n$th Landau level to 
the ground level with the creation of the neutrino-antineutrino pair takes the form: 

\beq
\label{eq:factor1_}
|{\cal M}_{e_n \to e_0 \nu \bar \nu} |^2 &=&  
\frac{G_{\mathrm{F}}^2}{n!} \, 
\exp{\left (-\frac{q^{\, \prime \, 2}_{\mprp}}{2\beta} \right )}
\left (\frac{q^{\, \prime \, 2}_{\mprp}}{2\beta} \right )^n \, 
\frac{M_n^2}{(p^{\, \prime} \tilde \Lambda q^{\, \prime})} \, \bigg \{ 
\left |C_V (p^{\, \prime} \tilde \Lambda j^{\, \prime}) - 
C_A  (p^{\, \prime} \tilde \varphi j^{\, \prime})\right |^2 
\\
\nonumber
&+& \frac{(j^{\, \prime} \tilde \Lambda j^{\, \prime \, *})}
{q^{\, \prime \, 2}_{\mprp}} 
\, \left [C_V (p^{\, \prime} \tilde \Lambda q^{\, \prime}) - 
C_A  (p^{\, \prime} \tilde \varphi q^{\, \prime})\right ]^2 -  
\frac{2}{q^{\, \prime \, 2}_{\mprp}} \, \left [C_V (p^{\, \prime} 
\tilde \Lambda q^{\, \prime}) - 
C_A  (p^{\, \prime} \tilde \varphi q^{\, \prime})\right ] 
\\
\nonumber
&\times&  Re \left ( (q^{\, \prime} \tilde \Lambda j^{\, \prime}) \,  
\left [C_V (p^{\, \prime} \tilde \Lambda j^{\, \prime \, *}) - 
C_A  (p^{\, \prime} \tilde \varphi j^{\, \prime \, *}) \right ] 
\right ) \bigg \}.
\eeq

With taking account of Eq.~(\ref{eq:weldon}), the amplitude squared of the process 
$\gamma e \to e \nu \bar \nu$ takes the form:
\beq
\label{eq:resamp2}
&&|{\cal M}_{\gamma e \to e \nu \bar \nu}|^2 
=  \sum\limits_{n=1}^{\infty} \; 
\int \frac{\dd^2 p''}{(2 \pi)^2 \, 2 E''_n} \, 
(2 \pi)^3 \, \delta^3 (P - p^{\, \prime \prime}) \, \frac{|{\cal R}_n|^2}{2 E''_n \, \Gamma_n} 
\\
\nonumber
&&= \sum\limits_{n=1}^{\infty} \; 
\int \frac{\dd^2 p''}{(2 \pi)^2 \, 2 E''_n} \, f_e (E''_n)\, 
(2 \pi)^3 \, \delta^3 (P - p^{\, \prime \prime}) \, \frac{|{\cal R}_n|^2}{2 E''_n \, 
\Gamma^{cr}_{e_0 \gamma \to e_n}}
\, .
\eeq
Here we have used the property of the $\delta$ function:
\beq
\delta (P^2_{\mprl} - m^2 - 2 \beta n) = \frac{1}{2 E''_n} \, \delta (P_0 - E''_n) \, , 
\eeq
where $E''_n = \sqrt{p^{\, \prime \prime 2}_z + m^2 + 2 \beta n}$.

Substituting Eq.~(\ref{eq:resamp2}) into the expression for the luminosity~(\ref{eq:Q}), 
and taking into account Eqs.~(\ref{eq:gammacr1}) and~(\ref{eq:factor}), we obtain:
\beq
\label{eq:Qtot}
Q_{\gamma e_0 \to e_0 \nu \bar \nu} = \sum\limits_{n=1}^{\infty}  Q_{e_n \to e_0 \nu \bar \nu} \, ,
\eeq
where 
\beq
\nonumber
Q_{e_n \to e_0 \nu \bar \nu} &=&  \frac{1}{L_x}\; \int 
\frac{\dd^2 p^{\, \prime \prime} }{(2\pi)^2 \, 
2 E_n^{\, \prime \prime}} \, f_e (E_n^{\, \prime \prime}) \,  
\frac{\dd^2 p'}{(2\pi)^2 \, 2 E^{\, \prime} } \, 
\left [1-f_e (E^{\, \prime}) \right ] \,  
\frac{\dd^3 p_1}{(2\pi)^3 \, 2 \varepsilon_1} \,
\frac{\dd^3 p_2}{(2\pi)^3 \, 2 \varepsilon_2} \,  
\\ [3mm]
\label{eq:Qnusynh}
&\times & q^{\, \prime}_0\,
(2 \pi)^3  \, \delta^3 (p'' - p^{\, \prime} - q^{\, \prime}) 
|{\cal M}_{e_n \to e_0 \nu \bar \nu}|^2 \, ,
\eeq
is the neutrino luminosity due to the process $e_n \to e_0 \nu \bar \nu$~\cite{Yakovlev2000}.


\section {Conclusion}\label{Sec:6}


In conclusion, let us summarize some of our results. 
 We have considered the neutrino photoproduction on an electron,
$e \gamma \to e \nu \bar \nu$, in dense magnetized medium
in both  resonance and non-resonance cases.
 The changes of the photon dispersion properties in a magnetized 
medium are investigated.
It has been  shown, that taking into account of the photon dispersion anisotropy 
in the limit of non-relativistic plasma leads to substantial modification 
of the neutrino emissivity 
due to the process $\gamma e \to e \nu \bar \nu$, if compared with the previously obtained results.
 We have obtained the most general expression for the neutrino emissivity due to the process
$\gamma e \to e \nu \bar \nu$  in relativistic and non-relativistic
plasma at an arbitrary relation between the plasma frequency and the temperature.
 It has been shown that the result known in the literature for 
the contribution of the photoneutrino process to the neutrino emissivity  
in the limit of a relativistic plasma was understated by several orders of magnitude. 
 It has been shown that in the case of resonance on the virtual electron,  the 
neutrino emissivity due to the process   $\gamma e_0 \to e_0 \nu \bar \nu$  can be  expressed 
in terms of the neutrino emissivity due to the process   $e_n \to e_0 \nu \bar \nu$.

\bigskip

{\bf Acknowledgements}  

This work was started together with Nickolay Vladimirovich Mikheev, who passed away on June 19, 2014. 
The paper is dedicated to the blessed memory of our teacher, colleague, and friend. 

\bigskip

We are grateful to  A.\,A.~Gvozdev, and I.\,S.~Ognev for useful discussions
and valuable remarks. A.\,K. and D.\,R. express their deep gratitude to the organizers of the 
Seminar ``Quarks-2014'' for warm hospitality.

\bigskip

The study was performed with the support by the Project No.~92 within the base part of the State Assignment 
for the Yaroslavl University Scientific Research, and was supported in part by the 
Russian Foundation for Basic Research (Project No.~\mbox{14-02-00233-a}).

\end{document}